# ABSTRACTION LEVEL TAXONOMY OF PROGRAMMING LANGUAGE FRAMEWORKS

Dr. Brijender Kahanwal

Department of Computer Science & Engineering, Galaxy Global Group of Institutions, Dinarpur, Ambala, Haryana, INDIA

## ABSTRACT

*The main purpose of this article is to describe the taxonomy of computer languages according to the levels of abstraction. There exists so many computer languages because of so many reasons like the evolution of better computer languages over the time; the socio-economic factors as the proprietary interests, commercial advantages; expressive power; ease of use of novice; orientation toward special purposes; orientation toward special hardware; and diverse ideas about most suitability. Moreover, the important common properties of most of these languages are discussed here. No programming language is designed in a vacuity, but it solves some specific kinds of problems. There is a different framework for each problem and best suitable framework for each problem. A single framework is not best for all types of problems. So, it is important to select vigilantly the frameworks supported by the language. The five generation of the computer programming languages are explored in this paper to some extent.*

## 1. INTRODUCTION

The language is the source of communication among human beings. Every country or region has its own language for the communication among their people, like Hindi, English, Urdu, French, etc. In the same sense the persons are required to communicate with the computer machines, so there is a need of such a language that can be understood by the computer. Such a language is known as a computer language or a programming language. Too many programming languages are presented to carry out varied kinds of jobs on the computer machine.

A set of rules and symbols are utilized by a computer programming language to operate on computer. We give command to the computer then it is converted in its own language (*Machine Language means 0's & 1's form instructions*). These languages can be classified in many ways. The programming languages are utilized to develop programs to work on computers.

The programmers develop the programs for computer. But in today's scenario the term programmer is very common and used for a simple computer operator. All the computer languages have their own set of rules and grammars. All these are well known as the syntax rules of any programming language. A developer always follows the syntax rules to write the correct code for the program or the application. Every day we heard of a new programming language means they are growing at a high speed, but only a few of them become popular among the users like C & C++.

                                                                 1

The very first language was made up of just two symbols 0 and 1 (*Machine language*). But in current time, we can make the computer programs using common English and Mathematical terms. The classification of the computer languages always vary according to the features selected. We can classify the languages, according to levels of abstraction which are as follows and it is also described with the help of Figure 1:

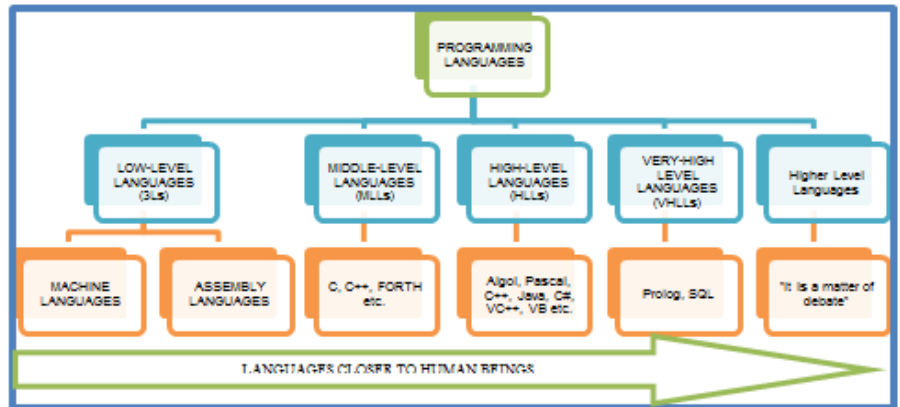

Figure 1. Abstraction Level Taxonomy of Programming Languages.

- ➢ Low-level Languages
  - ○ Machine Languages
  - ○ Assembly Languages
- ➢ Middle-level Language
- ➢ High-level Languages
- ➢ Very High-level Languages
- ➢ Higher level Languages

## 2. LOW-LEVEL LANGUAGES

Low-level languages are direct machine codes or very close to the machine. It provides a little or no abstraction from a computer's instruction set architecture. The word 'low' here refers to the small or nonexistent amount of abstraction between the language and machine language. Such types of languages are very close to the hardware.

There is no need of the translator such as compiler, interpreter for such type of languages. All the low-level programs are very quickly run able on computer machines and a very few memory space is needed as comparative to the high-level language programs. Such types of languages are simple, but are difficult to use because of so many technical details which must be memorized by the developers.

A computer can only understand and execute the instructions of the languages which are in the form of 0's and 1's (Binary number system). Low-level languages are machine oriented and always require the extensive knowledge of the computer architecture (computer hardware & computer configuration).

There are two types of low-level languages named machine languages and assembly languages.

## 2.1. Machine Languages

These are the lowest and most basic level of programming languages. It was the first type of computer language developed. The computer can understand only the special signals, which are the high level current and the low –level current and 1's and 0's represent them respectively. These two digits are called binary digits. Computer can understand only the programs which are written in binary digits. Such types of programs are called as Machine Language programs.
These are the only languages which are directly understood by the computers.  A series of codes is supplied to the computer then it identifies the fed code and converts them in to the electrical signals to run it as required. As an example:

| | | | | |
|---|---|---|---|---|
| 1010 | 0011 | 0001 | 1001 | (*Machine Language*) |
| ADD | R3, | R1, | R9 | (*Assembly Language*) |

The language, which uses binary digits, is called the machine level language. Machine language has its own pros and cons.

1. *First Generation Languages:* It was the first language to be programmed by the programmers. It is the first generation computer languages.
2. *Machine Dependent:* The internal design of every machine (architecture) is different from another machine, so the machine coding also differs. It says that the program instructions which are designed for one kind of machine cannot be utilized on another kind of machine.
3. *Fast processing:* All the machine instructions are directly recognized by the computing machine; the machine language programs are very quick to run.
4. *Error prone:* The instructions are written using 0's and 1's, so it is a very cumbersome   task. Hence, there are more chances of error prone codes in these languages.
5. *Difficult to use:* The binary digits (0 and 1) are used to represent the complete data or instructions, so it is a tough task to memorize all the machine codes for human beings.
6. *Difficult of debug:* When there is a mistake within the logic of the program then it is difficult to find out the error (bug) and debug the machine language program.
7. *Difficult to understand:* It is very difficult to understand the existing programs; it requires a great knowledge of machine code with the system architecture. For every machine the machine code is different. We cannot work easily on the other configuration system if we have the knowledge of one machine code.
8. *Efficient code for the Machine:* The coding of the machine language programs is very efficient for the machine.
9. *No need of Translator:* The machine code is directly understood by the machines, so no need of the translator.
10. *Programmer Requires the Knowledge of Computer Architecture:* To program in the machine languages, the programmer need to understand the computer architecture.
11. *Need to remember a lot of machine codes:* We have to remember a lot of machine codes for programming in these languages.
12. *Need to remember all memory addresses:* We have to remember all memory addresses for programming.
13. *Different Machine language for the different computer:* Every computer has its own machine language.

*Applications of Machine Languages:* It was used by the programmers when there was no computer language, but now many options are available for programming. The programming in machine languages needs a lot of time to learn and to document. So it's always avoided by the programmers. Assembly languages are used in place of them.

## 2.2. Assembly Languages

This is the next level of programming languages after the machine languages. In assembly languages, we use alphanumeric symbols (as operands and operations) instead of binary digits. These mnemonics can have up to five letter combinations maximum e.g. MOV, INC, SUB, ADD, MUL, START etc.

These languages are also known as "*Symbolic Programming Languages*" because of the usage of symbols to represent operation codes and storage locations. The programs written in assembly language are called assembly codes. Assembly language is easier to understand by the humans as compared to its predecessor language (*Machine Language*). The coding of the both of these languages is shown in the Table I.

Table I. The Machine Languages Code and the Corresponding Assembly Language code

| Language Code (Machine) (16-BIT INSTRUCTION SET) | Assembly Language Code (Equivalent) | | |
|---|---|---|---|
| 1000000100100101 | LOAD | R1 | 5 |
| 1000000101000101 | LOAD | R2 | 5 |
| 1010000100000110 | ADD | R0 | R1 R2 |
| 1000001000000110 | SAVE | R0 | 6 |
| 1111111111111111 | HALT | | |

The accountings engaged in the machine language programming are too much uninteresting. At the time of modification of the machine languages programs, always change the addresses of the data items. The developers will always take care of the whole program code and the bit pattern utilized by him/her.

It is human nature to be notorious at the easy tasks which are repeated very early. Assembly languages are more user-friendly. Machine language commands are replaced by *mnemonic* commands (*operation code*) on a one-to-one basis. The translator programs always changes the mnemonic code to its equivalent machine coding.

The symbolic addresses are utilized for the data times by the developers. The assembler programs allot the machine addresses and make certain that the different data items do not overlie in computer memory, an unenthusiastically ordinary incidence in machine language program codes. The assembly program coding is generally separated into diverse fields which split with spaces or tabs. A typical assembly code line is shown below:

    *[Label]     [Op-code]    [Operand1],    [Operand2]    ; Comment*

There are some important terms regarding Assembly Languages which are as follows:

1. *Labels* are the starting field of assembly language instructions. It may be blank. If it is there then the assembler program describes the label with corresponding to the address in which 1st byte of the machine code created for the specified statement will be stacked.
2. *Op-codes* are basically the Mnemonics. These are operation codes which are assigned to each processor instruction (set). The assemblers translate the op-codes into their binary equivalent code. Mnemonics are readable codes by us which are for the independent op-codes.
3. *Operands* are the objects which are used by the operation represented by the Op-code.
4. *Comments* are used after the semicolon to describe about the coding. These are ignored by the assemblers when they translate the assembly language program.
5. *Processor Instructions (set)* are classified in the following categories:

(a) *Data Transfer Instructions:* These instructions are used to transmit data from one position to another. The data transfer may be register-to-register, register-to-memory, memory-to-register, immediate value to register such as MOV, MOVX, PUSH, POP, etc.

Example: MOV         AX,   5                ; AX ← 0x05

(b) *Arithmetic Instructions:*     These are the instructions which are used to do the arithmetic operations like INC, ADD, MUL, SUB, DIV, DEC, etc.

Example: ADD AX,    BX            ; AX ← AX + BX
          MUL CX               ; AX ← AX * CX
          INC  AX              ; AX ← AX + 1
          DEC CX              ; CX ← CX – 1

(c) *Logical Instructions:* These are the instructions which are used to do the logical operations like AND, NOT, XOR, CLR, PROC, ORL, XRL, etc.

Example: AND AL,    5           ; AL ← AL & 00000101
          NOT  AL           ; AL ← ~ AL

(d) *Control Transfer Instructions:* Such instructions are jumping ones which are utilized to jump from one place to another in a program and utilized for controlling loops. The instructions are like JZ, JNZ, JE, JMP, AJMP, LJMP, ACALL, LCALL, RET, etc.

Example: JMP L12              ; jump to line 12
          JNZ  L12          ; jump to line 12 if not zero
          JZ    L12         ; jump to line 12 if zero

The first field is optional. It is the label field. It is used to specify symbolic labels and constants. The few assembler programs always need the labels with the colon as a delimiter ([Label :]). The next field is the op-code (*mnemonic*) field. The 3rd and the next fields are operands. These are usually comma-separated. The comments start with a delimiter semicolon (;) and ends in the same line.

| Operation code | Address |
|---|---|
| LOAD | A |
| ADD | B |
| STORE | C |

Clearly scope of error is very much reduced. However, the computer does not directly understand assembly languages. A program called an assembler translates it into machine language. The mnemonic op-code like ADD is replaced with the equivalent machine code by the assembler. And it allots storage addresses to every symbolic variable which are utilized by the developers. The assembler assigns the addresses to symbols A, B, and C and make sure that all the addresses are different from each symbol. So for making easy process for the programming, an additional process level has expanded.

In the real time applications the assembly language is still utilized for controlling the computer activities with efficiency. So the languages always require a sound knowledge of the machine architecture. For example, different ADD instructions are required for various kinds of data item. Assembly languages are machine-oriented so, the developers are always requiring re-writing the programs to implement that on the different computing machine.

The salient features of assembly languages are given below:

1. *2ⁿᵈ Generation Languages:* These are the second generation languages after machine languages.
2. *Easy to be compared with Machine languages:* The programs written in the Assembly language can be easily comparable with the machine language programs.
3. *Easy to understand:* As compared to the machine languages these are easier to understand.
4. *Easy to remove errors:* Because of the codes use English alphabets, it's easy to locate and correct errors in an assembly language programs.
5. *Easy to modify:* As the easy understanding of the programs of these languages, so we can modify them easily as the comparison with machine programs.
6. *Machine dependent:* A program which is written for one machine cannot be rerun of the different machine because every machine has a different architecture. So these are machine dependent.
7. *Need of Translator for the execution of the Program:* The Assembly language programs cannot be directly run or execute on the system, they always need the Assemblers. These tools convert the assembly programs into equivalent machine codes.
8. *Programmer Requires the Knowledge of Computer Architecture:* To program in the Assembly languages, the programmer need to understand the computer architecture as for as the machine languages.
9. *Efficient code for the Machine:* The coding of the Assembly language programs is very efficient for the machine.
10. *Fast processing:* As the assembly instructions are directly converted to the machine code by the assemblers and that is directly understood by the computer; so they are processed (executed) very quickly.
11. *Different Assembly Language for the different computer:* Every computer has its own set of instructions especially according to the computer architecture.
12. *Used for Specific Applications:* Used in applications which are cost sensitive (Washing Machines & Music Systems) & time critical (Aircraft Controls).
13. *Don't solve User's all Programming Problems:* For printing a number, find and respond to a specific command of teletypewriter, these tasks should be translated into a sequence of simple computer instructions by the assembly language programmer. Then they can be implemented in the assembly programming.
14. *Assembly Language Programs are not Portable:* As we know that every microcomputer has its own assembly language. So one assembly language program cannot be run on another microcomputer.

*Applications of Assembly Languages:* The languages are used by the programmers when they have to process the limited data, when the memory cost factor is accounted (*washing machines*), when the programs are very small, when there is a need of real-time applications (*aircraft control*), and when the applications requires more input or output or control computations.

## 3. MIDDLE –LEVEL LANGUAGES

Middle-level languages have been developed in recent years to bridge the gap of high-level and low-level languages. Some of these languages fall in the category of object-oriented e.g. C#, C++, Java, and FORTH. These languages are helpful in developing graphical user interfaces (GUIs) that run on personal computers. Middle-level language programmers need more technical skills as compared to the high-level language programmers. These languages are closely related to man and machine.

Such languages offer a small set of controlling and data-manipulating instructions which can be utilized by the developers. A high-level language is designed to give programmers all they could possibly want already in-built. A low-level programming every function is defined by the developers since nothing is predefined and libraries of them available.

These are the building block programming languages, because all the routines are created firstly for the specific functioning and make a collection or library of them. C and C++ languages allow a programmer to define routines to perform high-level commands. These routines are called functions. These are very important to C and C++. We can tailor a library of C and C++ functions to perform tasks that are carried out by our program. The overlapping of the middle level languages over the other levels is shown in the Figure 2.

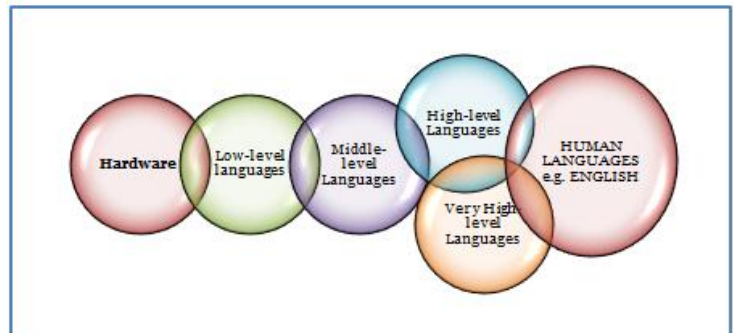

Figure 2. Overlapping of Middle-level languages on other levels (Low – level and High – level) of languages.

Main lines of the middle level languages are as they carry features like inline assembly language programs; accessing memory directly using pointers; use of system registers for fast processing; supporting high-level language features; user friendly nature; more close to machines; need of more effort to program; and these are far from human beings.

## 4. HIGH –LEVEL LANGUAGES

High – level languages are problem-oriented rather than computer-oriented. There are so many low-level language instructions corresponding to every high-level language instruction. Compilers and Interpreters are used to translate the high – level language source programs into machine language instructions.

To write large pieces of software, like web browsers, word processors, computer games, audio-video players, or a library system, cannot be done effectively by writing all of its instructions with the help of low-level programming languages. Such type of software is written in high-level programming languages. These languages allow the complex sequences of processor instructions to be expressed briefly.

High-level programming languages have formats close to English language. The main purpose of developing these languages is to write programs easily. These are basically symbolic languages. Here, English words and/or mathematical symbols are utilized in place of mnemonic codes.
There are so many high-level languages exist for different type of works. FORTRAN (FORmula TRANslation language) is the oldest high-level language which was used for expressing the algebraic notations. COBOL (COmmon Business-Oriented Language) is the language for business applications. C is a system programming language. A few languages are GUI (Graphical User Interface) based to develop applications.

Some of the features of high – level languages are given here:

1. *3rd Generation Languages:* This generation of programming languages is the refinement to the 2nd generation languages.
2. *Understandability:* The users can easily understand the programs written in these languages.
3. *Debugging:* We can easily find the errors (bugs) in programs written in these languages.
4. *Portability:* It is easy to run the same program which is written on one machine and run on the different machine.
5. *Easy to Use:* These languages are English like and we can write programs easily, if we are well know to English language.
6. *Problem-oriented Languages:* We can write the programs like the language of the problem. As there are business-oriented languages and as well as scientific languages.
7. *User-friendliness:* Because of the English like coding we are saying these are user-friendly. The users are very well aware about the human language "English".
8. *Little time to write Programs:* In these languages the writing of programs requires less time.
9. *Easy Maintenance:* We can easily maintain the programs written in these languages.
10. *Machine Independence:* The programs developed on a machine can be executed easily on another machine with distinct configuration.
11. *Need of a Translator:* There is always a need to translate the high-level language program to machine language program.
12. *Less Efficient:* The use of resources (*processor time/memory space*) of the programs written in these languages is high as compared to the low-level language programs. So, we are saying the programming is less efficient.

## 4.1 Taxonomy of High-Level Languages

Many languages have been developed for attaining diverse works, some are fairly specialized, and others are quite general purpose. The taxonomy of the high-level languages is shown in the following Figure 3.

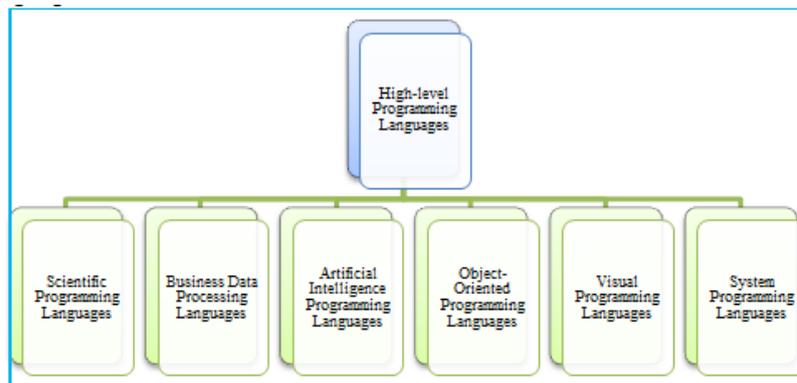

Figure 3. The taxonomy of high – level programming languages

These languages are briefly described here.

*a) Scientific Programming Languages:* These languages are oriented towards the computational procedures to solve mathematical and statistical problems. Some of the languages are discussed as follows:

*FORTRAN:* Formula Translation has a strong presence in Computational Physics and Chemistry in areas including climate modeling, fluid dynamics and molecular dynamics.

*C*: It is used in more environments than FORTRAN; Operating Systems are written; embedded devices programming; and many games and other applications are developed in C.

*MATLAB:* So many built in complex mathematical functions are offered by the language, like operations on matrices. So it is easy to do the functional programming on definite domains. Though, the MATLAB programming utilizes both the translators interpreter as well as the just in time compiler. So the programs written in it are much slower than the corresponding C and FORTRAN programs.

*Mathematica:* Basically it was constructed for the symbolic manipulations. But later it has emerged as complete programming language. The basic utilization of the language was for the mathematicians and the scientists, because of so many built in functions for many fields. Like the MATLAB, it also provides tools to visualize the problems. Though, it has the same drawbacks regarding the execution time of the programs as was of the MATLAB.

*FIDIL (FInite DIfference Language):* It is another functional as well as scientific programming language.  The more complex mathematical operations which will operate on the complete array elements rather a single element of the array can be easily implemented in it.

*Fortress (Sun Microsystems):* It improves programmer's productivity for scientific and engineering applications. A novel HPC (High Performance Computing) programming language proposed with the idea of high programmability. Mathematical syntax is offered for the scientists to do programming in their scientific language. It also provides support for easy and correct parallel programming.

*b)      Business Data Processing Languages:* These languages give stress to basically large data processing applications for the businesses e.g. COBOL (Common Business-oriented Languages), RPG (Report Program Generator) etc.

*c)      Artificial Intelligence:* The string manipulations with searching patterns, and insertion as well as deletion of characters e.g. IPL (Information Processing Language), LISP (List Processing), Prolog (Program in Logic) etc.

*d)      Object-Oriented Programming Language:* It is a framework for developing the applications according to the real world concepts. As we know that the world is a complete collection of objects and each object has its own identity, states, and behaviour. The same concept is taken in these languages. The main concepts of the OOP are the data abstraction, polymorphism, and inheritance. Object-orientation provides a new view of computation. A software system is a collection of objects as the world is. Every object cooperates with the other objects through message passing to solve the real world problems.

Some of the programming languages which come under this category are SIMULA, SMALLTALK, EIFFEL, C++, Java etc.

*e)      Visual programming language (VPL):* So many dimensions are utilized here to describe the semantics. Such type of added dimensions uses the multi-dimensional objects, the spatial associations are utilized, or the time based dimensions are utilized to describe the "before-after" semantic associations. All these multi-dimensional objects or associations are the tokens and the collection these tokens are the visually based expressions. The visual expressions utilized in these may have a diagram, a free-hand sketch, an icon, or revelation of actions achieved by graphical objects. Where the visual expressions are involved, those are called the visual programming languages. These are also know as the object-based programming languages. Some of the programming languages which use this concept are Visual Basic (VB), Visual Java, and Visual C++ etc.

*f)*      *System Programming Languages:* These are the programming languages in which all the machine instructions can be written; the complete control on storage allocation and management is done; the operating system can be written; and parallel processing (synchronization) can be achieved. Some of the languages among these are CPL (Computer Programming Language), BCPL (Basic Combined Programming Language), C, and C++ etc.

*Applications of High-level Languages:*  These are the languages which are used by most of the programmers. As we know that these are very near to human beings. It means that close to the natural languages like English. These languages are used when the programs are very long, when the applications are of low-volume, when there are so many modifications required in the programs, when the applications need so much computations other than the input or output or control computations, when the memory requirement by the application is very large, and when compatibility is needed with the similar applications.

# 5. VERY HIGH –LEVEL PROGRAMMING LANGUAGES

Most of the fourth generation languages (4GLs) are non procedural languages. Here only the encouragement to the users and the developers to describe the results they need. Though, the series of instructions are decided by the computing machines which will provide the required results. Thus, 4GLs have helped in simplifying the programming process.  The Normal languages are 4GLs which are near to the English and other languages of human beings.

A 4GL is used in six major areas. Those areas include, Data input; Data management; Data analysts; Data output (including reporting); Graphics; user oriented interfacing only that includes the utilization of windows, and predefined screens etc. One example is shown as follows.

Multiply the numbers A and B

And put the result into C

All the forth generation programming languages are so much interactive and a dialogue in between the human being and the computer machine is supported. They are powerful system software because they provide default data management, programming, and reporting functions etc.  Consequently, this leads to Decreased Development Time; Decreased Development Costs; Increased Software Quality; Improved Decision Making Capabilities; and Increased Availability of Information.

As an example, the Oracle is a 4GE (4[th] Generation Environment) which has the components like End-user Query Language (e.g. SQL), Screen Formatter (e.g. Oracle's screen painter in SQL *Forms), Report Generator (SQL *Report), Data Dictionary (SQL *DD), SQL *Plus, SQL *Forms, SQL *Graph and many more. Other 4GLs are Progress 4GL, AVS, APE, LINC, BuildProfessional, GEMBase, and Mathematica etc.

4GLs can be classified in five major groups: Programming extensions to the operating system through command interpreters; database management and query; new programming languages such as graphics; productivity-oriented tools, through pre-compilers; and spreadsheet systems and integrated software.

All these languages have their own advantages as well as disadvantages like other computer languages. Only the user asks for the required tasks and the corresponding programming code is generated directly from the specifications.

## 6. HIGHER –LEVEL PROGRAMMING LANGUAGES

This class of programming languages comes under the category of fifth Generation Languages (5GLs). These are yet to come. These are the subjects of discussion in the programming community. These will use the ideas of Artificial Intelligence (AI) to cause novel software to exists. Therefore, it is very characteristic job to develop such languages. There will be no significance of the algorithms. Only the constraints will be thought carefully in the programming. Nowadays Fifth Generation systems (5GSs) characteristically have large scale parallel processing (many instructions being executed simultaneously), different memory organizations, and novel hardware operations predominantly designed for symbol manipulation. The concept of parallel processing has come in place of the single central processor. This new hardware organization is often fastened with software that concentrates on representation of knowledge.

Though, the developers are making effort with these $5^{th}$ Generation computing machines and are utilizing HLLs (e. g. Prolog 2). It is well suited to make reservation of the term $5^{th}$ GL for the languages yet to come where the users will communicate with the systems to do programming. In these languages, we must design interface between human being and machine to permit affective use of natural language and images. In this human verbal point of view, this class of programming language will be the last set of the language generations. Ultimately, the computer will directly understand human beings.

The 5GLs have some benefits like the encouragement to initiate for the unskilled users; the encouragement of the upper management for the computers use; less time needed to obtain knowledge of the complex systems with much reduced annoyance; the use of computers for the any kind of community will increase.

These fifth generation languages will have their own negative aspects like lack of precision; not good in expressing the precise and complex logics; unable to express the tidy and smart structures; semantic overrun can be encouraged; will take much time to key in sentences; ambiguousness and confusion will be augmented; and also these languages will increase requirement of much processing.

## 7. CONCLUSION

This taxonomy has attempted to recognize and explain the salient characteristics of the programming languages. The five generations of computer languages are illustrated here. The fifth generation has been up till now to approach and it is a matter of argue. In addition, the taxonomy of the high-level languages is also portrayed which also presents light on the some of the main languages. Such taxonomy doesn't take into consideration all the possible features of all the languages. We also depict some of the pros and cons of these generations of the languages. On the other hand, it is expected that the article will be beneficial for the researchers as well as the computer language learners.